\documentclass[secnumarabic,showpacs,amsmath,amssymb,floatfix,superscriptaddress,nofootinbib, nobibnotes, aps, prl, twocolumn]{revtex4-2}

\usepackage{graphicx}

\usepackage{epstopdf}
\usepackage{color}         			% color text
\usepackage[normalem]{ulem}			% Allow strikeout formatting of text
\usepackage{soul} 						 	% strikeout formatting
\usepackage{lineno}
\usepackage{multibib}

\begin{document}
%\psdraft

\title{Slow Equilibrium Relaxation in a Chiral Magnet Mediated by Topological Defects}

%\date{\today}

\author{Chenhao Zhang}
\altaffiliation{These authors contributed equally to this work}
\affiliation{School of Physical Science and Technology, ShanghaiTech University, Shanghai 201210, China}

\author{Yang Wu}
\altaffiliation{These authors contributed equally to this work}
\affiliation{School of Physical Science and Technology, ShanghaiTech University, Shanghai 201210, China}

\author{Jingyi Chen} 
\altaffiliation{These authors contributed equally to this work}
\affiliation{School of Physical Science and Technology, ShanghaiTech University, Shanghai 201210, China}

\author{Haonan Jin} 
\affiliation{School of Physical Science and Technology, ShanghaiTech University, Shanghai 201210, China}
\affiliation{ShanghaiTech Laboratory for Topological Physics, ShanghaiTech University, Shanghai 201210, China}

\author{Jinghui Wang} 
\affiliation{School of Physical Science and Technology, ShanghaiTech University, Shanghai 201210, China}
\affiliation{ShanghaiTech Laboratory for Topological Physics, ShanghaiTech University, Shanghai 201210, China}

\author{Raymond Fan}
\affiliation{Diamond Light Source, Harwell Science and Innovation Campus, Didcot OX11~0DE, United Kingdom}

\author{Paul Steadman}
\affiliation{Diamond Light Source, Harwell Science and Innovation Campus, Didcot OX11~0DE, United Kingdom}

\author{Gerrit \surname{van der Laan}}
\affiliation{Diamond Light Source, Harwell Science and Innovation Campus, Didcot OX11~0DE, United Kingdom}
% \email{Gerrit.vanderLaan@diamond.ac.uk}

\author{Thorsten Hesjedal}
\affiliation{Department of Physics, Clarendon Laboratory, University of Oxford, Oxford OX1~3PU, United Kingdom}
% \email{Thorsten.Hesjedal@physics.ox.ac.uk}

\author{Shilei Zhang$^{\dagger}$}
\affiliation{School of Physical Science and Technology, ShanghaiTech University, Shanghai 201210, China}
\affiliation{ShanghaiTech Laboratory for Topological Physics, ShanghaiTech University, Shanghai 201210, China}
\affiliation{Center for Transformative Science, ShanghaiTech University,  Shanghai 201210, China }
%\email{shilei.zhang@shanghaitech.edu.cn}

%%%%%%%%%%%%%%%%%%%%%%%%%%%%%%%%%%%%%%%%%%%%%%%%%%%%%%%%%%
% Abstract 
%%%%%%%%%%%%%%%%%%%%%%%%%%%%%%%%%%%%%%%%%%%%%%%%%%%%%%%%%

\begin{abstract}
%
%Non-collinear spin textures, such as helices and skyrmions are important magnetic phases for the study of topological magnetism. 
%
%They usually present as equilibrium states, hence the stabilisation process is expected at a timescale of nanoseconds. 
%
We performed a pump-probe experiment on the chiral magnet Cu$_2$OSeO$_3$ to study the relaxation dynamics of its non-collinear magnetic orders, 
employing a millisecond magnetic field pulse as the pump and  resonant elastic x-ray scattering as the probe.  
Our findings reveal that the system requires $\sim$0.2\,s to stabilize after the perturbation applied to both the conical and skyrmion lattice phase; significantly slower than the typical nanosecond timescale observed in micromagnetics. 
This prolonged relaxation is attributed to the formation and slow dissipation of local topological defects, such as emergent monopoles. 
By unveiling the experimental lifetime of these emergent singularities in a non-collinear magnetic system, our study highlights a universal relaxation mechanism in solitonic textures within the slow dynamics regime, offering new insights into topological physics and advanced information storage solutions.
\end{abstract}

%%%%%%%%%%%%%%%%%%%%%%%%%%%%%%%%%%%%%%%%%%%%%%%%%%%%%%%%%
%introduction

\maketitle
Non-collinear spin textures that exhibit incommensurate long-wavelength modulations are important types of magnetic orders that provide a unique playground for the study of topological physics \cite{Pf_MnSi_Science_09, Tokura_FeCoSi_LTEM_Nature_10, Togawa_CrNb3S6_LTEM_PRL_12, Tokura_review_emergent_12, Tokura_review_skyrmion_Natnano_13,  Tokura_GdRuSi_REXS_Science_19, Tokura_GdRuSi_review_Advmater_22}.
They are usually formed as solitonic structures \cite{Bogdanov_JETP_89}, such as one-dimensional (1D) helices \cite{Tokura_FeCoSi_LTEM_Science_06}, two-dimensional (2D) skyrmions \cite{Pf_sim_Rob_Bog_06_Nature,Pf_MnSi_Science_09, Tokura_FeCoSi_LTEM_Nature_10}, or three-dimensional (3D) topological textures \cite{Rybakov_chiral_bobber_PRL_15, Donnelly_GdCo2_hard_vector_tomo_ptycho_Nature_17, Du_FeGe_EH_bobber_Natnano_18, Kiselev_toron_PRB_20,  Hatton_FeGe_STXM_Natcommun_20, Bobber_MCTR_PRL_21, CoTb_PRL_21, Du_FeGe_LTEM_bundle_Natnano_21, Du_braid_Natcommun_21, Fert_CoAl_MFM_cocoon_Natcommun_22, Tokura_Hard_tomo_STXM_Natmater_22, Infinity_Rod_Nanolett_23, Sideface_Screws_PRB_23, Kiselev_FeGe_LTEM_hopfion_Nature_23}. 
Their unique topology induces emergent phenomena \cite{Tokura_review_skyrmion_Natnano_13, Tokura_review_emergent_12}, and offers advanced information storage solutions \cite{Fert_sim_Natnano_13, Jiang_blowing_bubble_Science_15, Fert_TaCoIrPtn_MFM_transport_Natnano_18}. 
Therefore, the time-resolved experimental characterization of their dynamic relaxation processes is of great significance for topological magnetism and spintronics.  
Chiral magnets are materials that host 1D, 2D, and 3D solitonic magnetic phases \cite{Pf_MnSi_Science_09, Bauer_book_16}. 
The presence of the Dzyaloshinskii-Moriya interaction (DMI) leads to an energetic nonlinearity, such that 1D spin spirals become the ground state \cite{Bogdanov_review_11}.
These helical modulations transform into the conical phase at elevated fields, while the skyrmion lattice (SkX) phase emerges in a narrow pocket of the phase diagram near the transition temperature $T_\mathrm{C}$ at finite field \cite{Bauer_book_16}. 
The magnetic phase diagram for Cu$_2$OSeO$_3$, shown in Fig.\ 1(d), is a rather universal representation of the equilibrium states of a typical chiral magnet \cite{Bauer_book_16}.

The typical timescales over which these complex spin textures order are governed by micromagnetics, where the eigenenergy of the underdamped system corresponds to a few GHz \cite{Tokura_CuOSeO_FMR_PRL_12, Pf_universal_dynamics_Natmater_15}. 
Consequently, it is generally expected that non-collinear magnetic phases will return to their equilibrium state within nanoseconds following an excitation \cite{Nagaosa_Sim_Skrymion_Gen_Natnano_13, Tokura_CoZnMn_PP_LTEM_Sciadv_21, Loosdrecht_GaV4S8_PP_MOKE_Natcomms_22, Pang_sim_ip_field_removal_PRB_22}, as illustrated in Fig.\ 1(a).   
This behavior has indeed been observed in spin-wave spectroscopy experiments \cite{Tokura_CuOSeO_FMR_PRL_12, Pf_universal_dynamics_Natmater_15}.
However, recent discoveries of 3D topological defects have highlighted the significance of slower timescales \cite{Rybakov_chiral_bobber_PRL_15, Rocsh_FeCoSi_monopole,Rosch_MAP_tension_PRB_14,Pf_FeCoSi_LTEM_SciAdv_17,Kiselev_FeGe_LTEM_Natphys_18,Du_FeGe_EH_bobber_Natnano_18,Hatton_FeGe_STXM_Natcommun_20,Zhou_CoZnMn_LTEM_Nanolett_20, Kiselev_toron_PRB_20, Tokura_FeGe_in_plane_LTEM_NL_20, Hatton_Cu2OSeO3_Zn_ac_metastable_CommPhys_21,Sabri_sim_Cu2OSeO3_defects_PRB_23}.
For instance, monopole/anti-monopole singularities emerging in chiral magnets during the phase-stabilization process manifest over considerably longer periods \cite{Rocsh_FeCoSi_monopole, Pf_FeCoSi_LTEM_SciAdv_17, Hatton_FeGe_STXM_Natcommun_20, Tokura_FeGe_in_plane_LTEM_NL_20, Tokura_Hard_tomo_STXM_Natmater_22}. 
These defects can appear as joint structures connecting two skyrmion strings \cite{Rocsh_FeCoSi_monopole,Rosch_MAP_tension_PRB_14,Tokura_Hard_tomo_STXM_Natmater_22} or as Bloch points terminating chiral bobbers \cite{Rybakov_chiral_bobber_PRL_15, Du_FeGe_EH_bobber_Natnano_18, Bobber_MCTR_PRL_21,Infinity_Rod_Nanolett_23}. 
In either case, the system inherently strives to expel such 3D defects due to their unrelaxed local energy density \cite{Rocsh_FeCoSi_monopole, Rosch_MAP_tension_PRB_14, Rybakov_chiral_bobber_PRL_15, Tokura_MnSi_quench_Natphys_16, Pf_FeCoSi_LTEM_SciAdv_17, Hatton_Cu2OSeO3_Zn_ac_metastable_CommPhys_21, Sabri_sim_Cu2OSeO3_defects_PRB_23},
resulting in extended timescales required to unwind these topologically entangled spin textures \cite{Tokura_MnSi_quench_Natphys_16, Pf_FeCoSi_LTEM_SciAdv_17, Haton_Cu2OSeO3_Zn_ac_metastable_PRB_19, Blugel_films_lifetime_PRL_20, Hatton_Cu2OSeO3_Zn_ac_metastable_CommPhys_21}. 

\begin{figure}[ht!]
\includegraphics[trim = 0cm 0cm 0cm 0cm, clip=true, width=8.5cm]{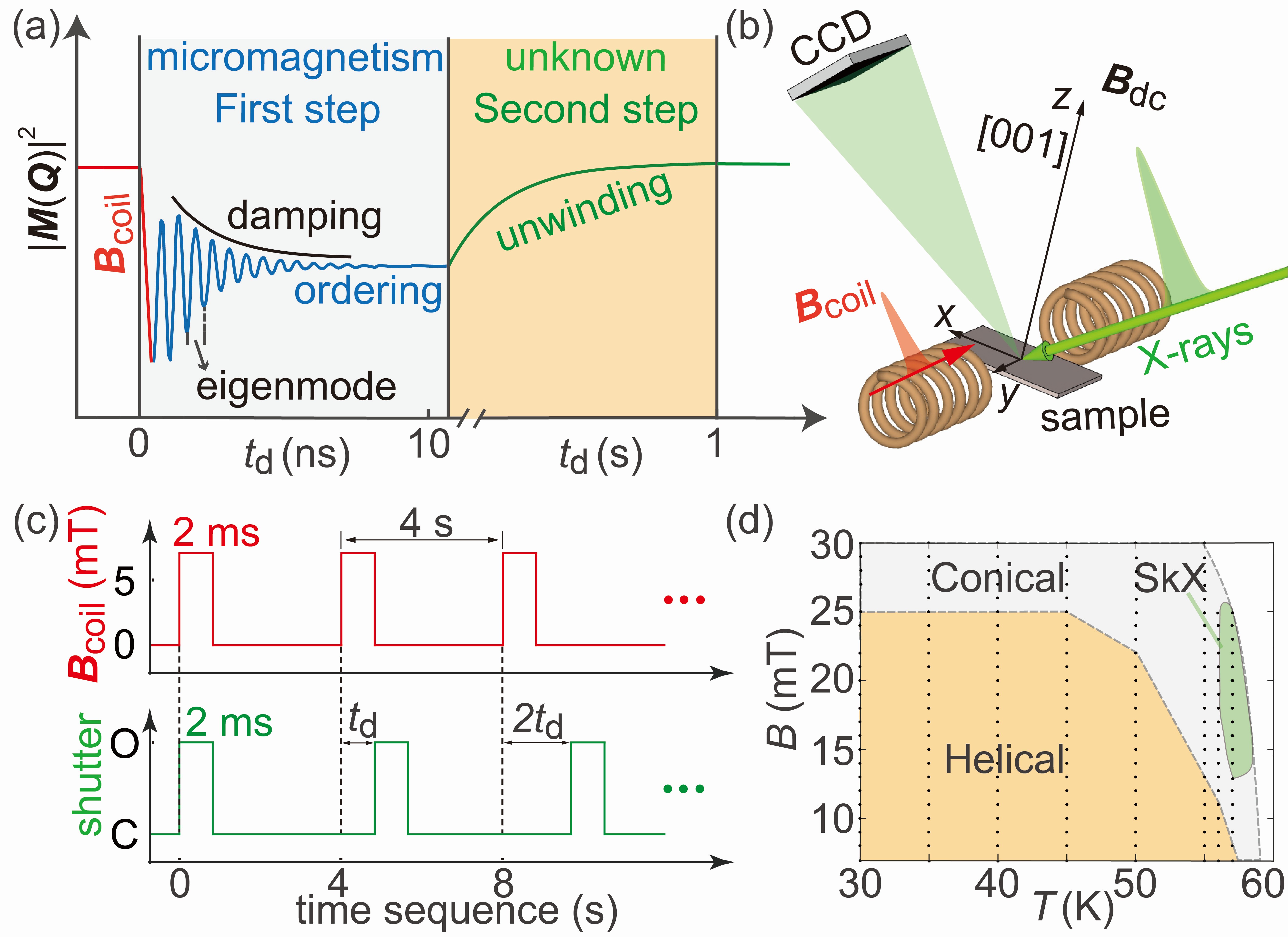}
\caption{(a) The timescales for a non-collinear magnetic order, at which various magnetic dynamics take place after a pumping field is applied. 
(b) Sketch of the scattering geometry, where an additional in-plane field pulse $B_\text{coil}=7$~mT is applied along the $y$-direction as the pump signal. 
(c) Time sequence of the measurement, illustrating the time delay $t_\text{d}$. 
(d) Equilibrium phase diagram of the measured Cu$_2$OSeO$_3$ single crystal, mapped via their signature REXS patterns ($B_\text{coil}$ switched off). 
}
\label{fig_1}
\end{figure} 

The lowest energy state is described by a real-space magnetic structure $\mathbf{m(r)}$, which is the superposition of each populated modulation $\mathbf{m(r)}= \sum_{Q}\mathbf{M(Q)}e^{2\pi i \mathbf{Q}\cdot\mathbf{r}}$ \cite{Pf_MnSi_Science_09, Rosch_3D_Monte_Carlo_PRB_13}, where $\mathbf{Q}$ are the modulation wavevectors and $\mathbf{M(Q)}$ the structure factor of the ordered magnetic crystal at a reciprocal lattice point $\mathbf{Q}$.
Although an ordered lattice is formed, it is speculated that local 3D topological defects may nucleate throughout the system \cite{Rosch_MAP_tension_PRB_14,Pf_FeCoSi_LTEM_SciAdv_17}. 
Unwinding these singular points requires additional time, with the system eventually reaching the global energy minimum.   
It is crucial to emphasize that such singularity-mediated equilibrium relaxation dynamics fundamentally differs from the previously reported metastability evolution process \cite{Pf_FeCoSi_SANS_PRB_10,Rocsh_FeCoSi_monopole,Pf_FeCoSi_ac_metastable_PRB_16,Tokura_MnSi_quench_Natphys_16,Pf_FeCoSi_LTEM_SciAdv_17,Tokura_FeGe_LTEM_LT_Natphys_18,Haton_Cu2OSeO3_Zn_ac_metastable_PRB_19,Hatton_Cu2OSeO3_Zn_ac_metastable_CommPhys_21},
which describe the conventional decay from a higher-energy metastable state to the ground state over a timescale of minutes. 
In our scenario, we propose a ubiquitous two-step relaxation process for non-collinear magnetic phases: a rapid initial relaxation to a stable ordered ground state within nanoseconds [Fig.\ 1(a)], followed by a slower dissipation of local topological defects, which occurs over seconds.
This extended dissipation time underscores the critical influence of topological defects on the dynamic behavior of these magnetic systems, not only being of importance for their fundamental understanding, but also profoundly affecting envisioned high-speed skyrmion device applications.

To test this hypothesis, we designed a pump-probe experiment to measure the time-resolved relaxation process of the soliton phases at slow timescales.  
Figure 1(b) shows the experimental setup based on resonant elastic x-ray scattering (REXS).
REXS is an effective probe for the characterization of the magnetic structure factor $\mathbf{M(Q)}$ of an incommensurate spin lattice phase \cite{Cu2OSeO3_REXS_PRB_16}.  
In REXS, the time resolution is limited by the setup and governed by an upstream mechanical shutter for the charge-coupled device (CCD) camera, which is typically chosen to be 2\,ms. 
A static field $\mathbf{B}_\text{dc}$ is applied in order to stabilize the magnetic phase of interest.  
It can be applied along the [001] crystalline axis (along the $z$-direction) to stabilize an out-of-plane (OOP) SkX phase, or it can be oriented along the $x$-direction to reach the conical and in-plane (IP) SkX order \cite{Cu2OSeO3_in_plane_REXS_NL_20}. 
Moreover, a pair of Helmholtz coils is added with their axis along the $y$-direction, through which an orthogonal magnetic field $\mathbf{B}_\text{coil}$ can be applied. 
Depending on the current-input mode, $B_\text{coil}$ can either be a field pulse (for a short duration of 2\,ms), or a static field to probe the static `excited' state resulting from the tilted field vector $(\mathbf{B}_\textbf{coil} + \mathbf{B}_\textbf{dc})$.
The pulsed field thus perturbs the (stabilized) magnetic order after which the time-dependent relaxation process is measured. 
More details on the experimental setup can be found in the Supplementary Materials \cite{SM}.

The timing structure is controlled by a built-in field-programmable gate array board, which triggers both the pulsed field $B_\text{coil}$ and the CCD shutter with controlled time delays and the duration.   
Figure 1(c) shows the time sequence of our measurement. 
Each pump-probe cycle takes 4\,s (2\,s of measurement and 2\,s of waiting time) \cite{SM}, which proves to be sufficiently long for the magnetic peaks to be stable after each excitation by $B_\text{coil}$.  
A square-shaped field pulse with a peak amplitude of 7\,mT and duration of 2\,ms is applied as the pump.
The time-zero reference is defined as the rising edge of each $B_\text{coil}$ pulse.
Consequently, the CCD shutter is open for an exposure time of 2\,ms at a controlled time delay $t_\text{d}$ after each current pulse.

Figure 1(d) shows the equilibrium phase diagram of the Cu$_2$OSeO$_3$ single crystal for $\mathbf{B}_\text{dc}\parallel$~[001] geometry with $B_\text{coil} = 0$.   
The helical, conical, and SkX states can be easily identified via their signature REXS patterns \cite{Cu2OSeO3_REXS_PRB_16,Cu2OSeO3_REXS_NL_2016}. 
Note that the IP phase diagram is very similar to the OOP one, however, the details such as the critical fields can differ due to the demagnetization effects for which we have not corrected \cite{Pf_MnSi_ac_SQUID, SM}.
It is worth emphasizing that the metastable skyrmion phase was not observed in our sample due to the applied field-cooling protocol, possibly resulting from the slow cooling rate applied here ($<$4~K/min) \cite{Pf_FeCoSi_SANS_PRB_10,Pf_FeCoSi_ac_metastable_PRB_16,Tokura_MnSi_quench_Natphys_16}.  
Our investigation focuses on three distinct relaxation processes, namely the \emph{conical relaxation} process, and \emph{SkX relaxation} processes in the OOP and IP geometries, respectively.   
As shown in Fig.\ 2, the system is first initialized into the stable state of interest under certain $(T, B_\text{dc})$ conditions (left column). 
Upon applying $B_\text{coil}$ for a duration of time, it enters into an excited state with a varied spin texture (right column). 
Once $B_\text{coil}$ is removed, the system returns to its initial equilibrium state again (left column), of which the decaying dynamics is essentially probed by the pump-probe measurement.   

\begin{figure*}[t!]
\includegraphics[trim = 0cm 0cm 0cm 0cm, clip=true, width=16.5cm]{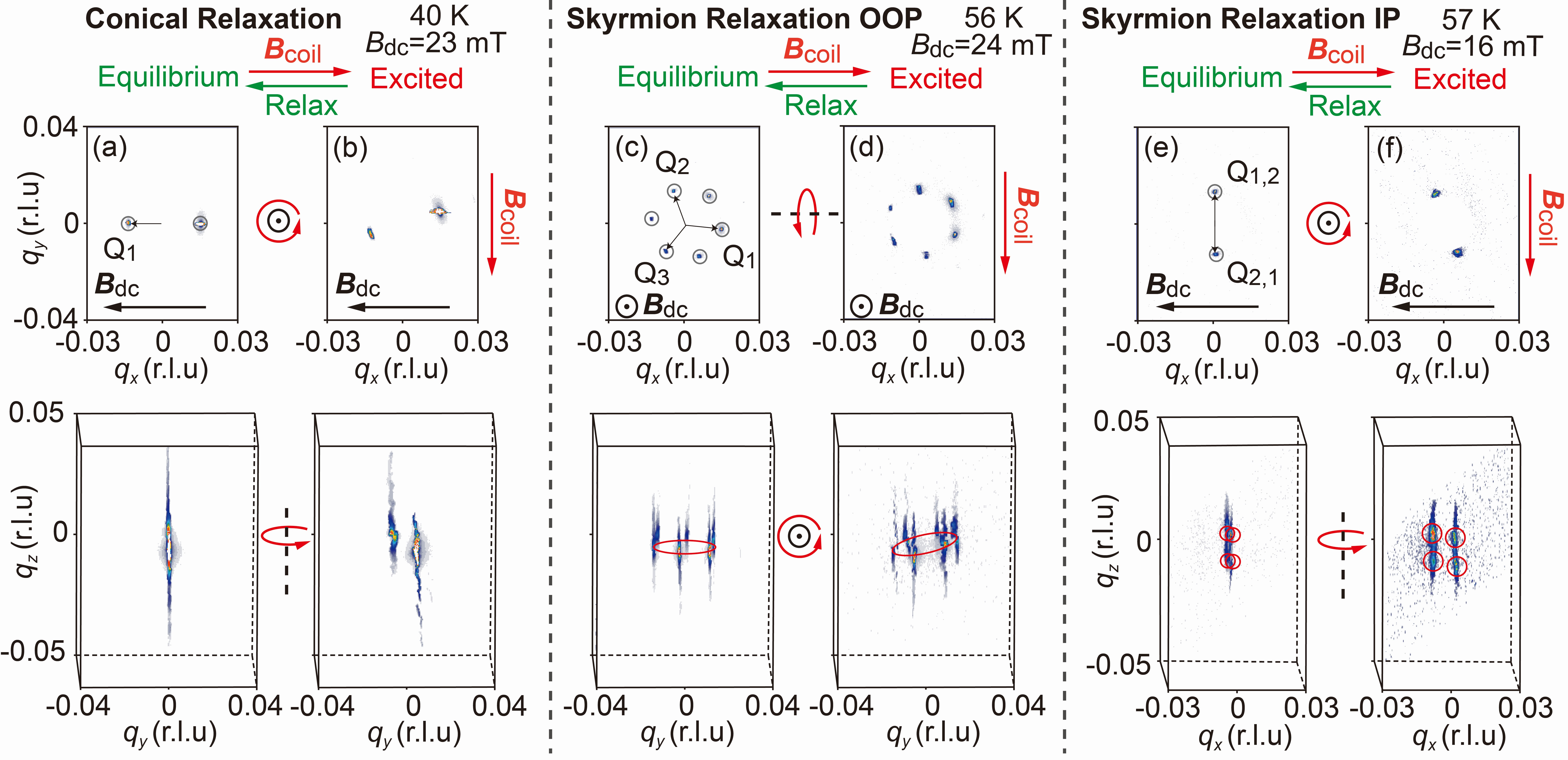}
\caption{Top view (top panels) and side view (bottom panels) of the 3D RSM representations on the static states for the three studied relaxation processes. 
Within each box, the left column shows the equilibrium state reciprocal space structures, while the right column shows the excited states after a static $B_\text{coil}=7$~mT is switched on. 
As indicated by the rotation symbols in-between the panels, the RSM pattern rotates around the $q_z$-axis for (a-b) and (e-f); while it rotates around the $q_x$-axis for (c-d).
The circles in (a,c,e) indicate areas with radius 0.003\,r.l.u.\ over which the intensity was integrated for the further analysis. 
}
\label{fig_2}
\end{figure*}

Figure 2 shows the 3D reciprocal space mapping (RSM) \cite{Bobber_MCTR_PRL_21, Infinity_Rod_Nanolett_23} of the static magnetic structures for the start and end points of the three dynamical processes, i.e., the equilibrium ($B_\text{coil}$ switched off) and the static excited states (constant $B_\text{coil}=7$~mT).  
As shown in Fig.\ 2(a), the conical state is initially prepared by an $x$-oriented $B_\text{dc}$ field at 40\,K. 
The spin spiral modulation is found within the $q_x$-$q_y$ plane at $q_z=0$, orientated along the $B_\text{dc}$ direction, featured by the $\mathbf{Q}_1$-peaks (Friedel pair). 
Moreover, as shown in the bottom panel of Fig.\ 2(a), the two $\mathbf{Q}_1$ peaks are elongated along $q_z$. 
This peak broadening is due to the finite penetration depth of the soft x-rays at resonance \cite{Bobber_MCTR_PRL_21, Infinity_Rod_Nanolett_23}. 
By switching on a static orthogonal $B_\text{coil}$ field of 7\,mT along the $y$-direction, the summed field vector $(\mathbf{B}_\textbf{coil} + \mathbf{B}_\textbf{dc})$ drives the conical peaks to an oblique angle. 
As shown in Fig.\ 2(b), the entire pattern undergoes a rotation of $18^\circ$ around the $q_z$-axis, suggesting that the extra static $B_\text{coil}$ field excites the system to a different lattice configuration. 
It is worth noting that although the pump source rotated the spin lattice by an angle, the system remains to be in the conical phase without encountering a phase transition. 
On the other hand, by removing the static $B_\text{coil}$, the system immediately recovers to the equilibrium state in Fig.\ 2(a).

Similarly, for the OOP SkX phase, the addition of the IP perturbing static field $B_\text{coil}$ essentially tilts the skyrmion plane by an angle of $15^\circ$ around the $q_x$-axis, as marked by the red loops in Fig.\ 2(c,d). 
The removal of $B_\text{coil}$ instantaneously sets the system back to the initial, unrotated SkX phase [as in Fig.\ 2(c)].  
Furthermore, the SkX in the IP geometry follows the same equilibrium-excitation-equilibrium behavior. 
As shown in Fig.\ 2(e), by applying a static $B_\text{dc}$ along $x$, the skyrmion plane lies within the $yz$-plane, leaving two pairs of magnetic ordering peaks $\mathbf{Q}_1$ and $\mathbf{Q}_2$ observable \cite{SM}, marked by the red circles in Fig.\ 2(e). 
The application of a static $B_\text{coil}$ subsequently rocks the skyrmion plane about the $q_z$-axis by $21^\circ$, as shown in Fig.\ 2(f), after which the system returns to the state of Fig.\ 2(e).  

\begin{figure}[t!]
	\includegraphics[trim = 0cm 0cm 0cm 0cm, clip=true, width=8cm]{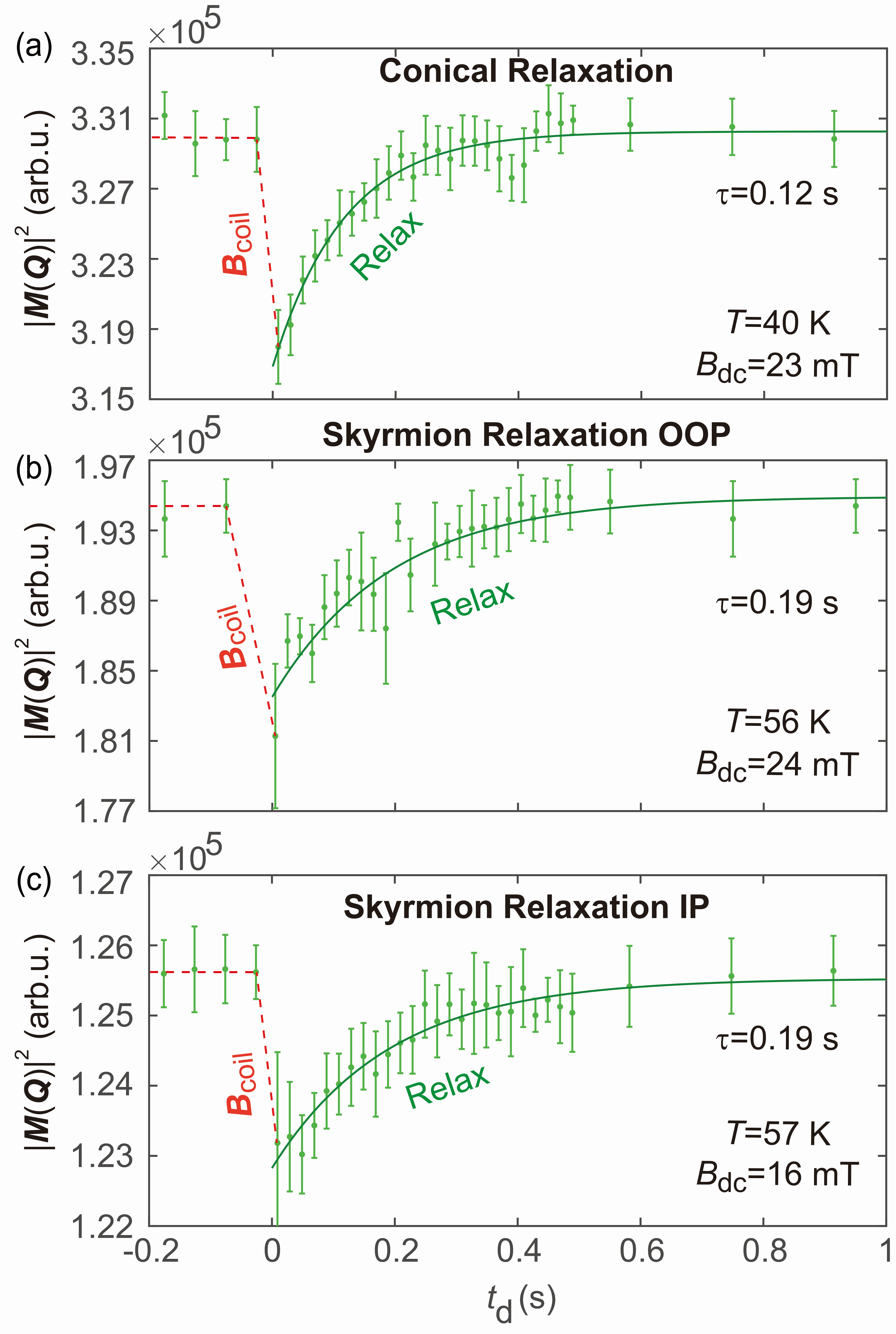}
	\caption{Pump-probe results for the three studied relaxation processes, showing the average over all magnetic peak intensities as a function of time delay. 
		The green dots are measured data points with error bars, and the green lines are the exponential fitting curve.
	}
	\label{fig_3}
\end{figure}

Having established the static magnetic structures, time-resolved pump-probe measurements were carried out to investigate their relaxation dynamics.
Now, as the pump, dynamic $B_\text{coil}$ field pulses of 7~\,mT were applied for a duration of 2\,ms.   
Figure 3 shows the magnetic peak intensity ($|\mathbf{M(Q)}|^2$) as a function of time delay for the three studied processes described in Fig.\ 2. 
While it should be expected that nonuniform magnetic phases will reach equilibrium on a nanosecond timescale \cite{Pf_FeCoSi_damping, Tokura_Cu2OSeO3_REXS_pumpprobe_PRL_17}, i.e., stationary peaks with constant intensity, we observed significantly prolonged relaxation processes,
reflected in the gradual change in peak intensity, towards equilibrium, i.e., as slow as $\sim$0.2\,s. 
Although the three processes correspond to fundamentally different magnetic phases and distinct relaxation paths, it was found that the slow dynamics of these magnetic orders show a universal behavior. 
For example, for the conical relaxation process [Fig.\ 3(a)], the peak intensity undergoes a sudden drop at $t_d=0$ by $4.5\%$, suggesting that the conical lattice become less ordered than the initial state at $t_d<0$.  
Despite the change in peak amplitude, it is important to note that the peak positions have no $t_d$ dependence at all. 
In other words, during the pump-probe measurement, the RSM pattern always remains the same as that shown in Fig.\ 2(a), and only the peak intensity evolves over time.   
The same behavior is also observed for the two skyrmion relaxation processes in Fig.\ 3(b,c).  
We thus conclude that the time-resolved measurements essentially probe the post-ordering evolution.
As the timescale that is required to stabilize a rigid lattice of a non-collinearly ordered state can be as fast as nanoseconds \cite{Nagaosa_Sim_Skrymion_Gen_Natnano_13, Tokura_CoZnMn_PP_LTEM_Sciadv_21, Loosdrecht_GaV4S8_PP_MOKE_Natcomms_22, Pang_sim_ip_field_removal_PRB_22}, our time resolution (limited to 2\,ms) is not fast enough to capture any details of the formation process of the $\mathbf{Q}$-wavevectors. 
Therefore, at $t_d = 0$, the excited state has already relaxed back to the ordered equilibrium state with well-defined $\mathbf{Q}$-configurations. 
The timescale in our case hence investigates the structure factor modifications, i.e., how the magnetic crystal becomes more ordered by dissipating its local defects.   

\begin{figure}[t!]
\includegraphics[trim = 0cm 0cm 0cm 0cm, clip=true, width=8cm]{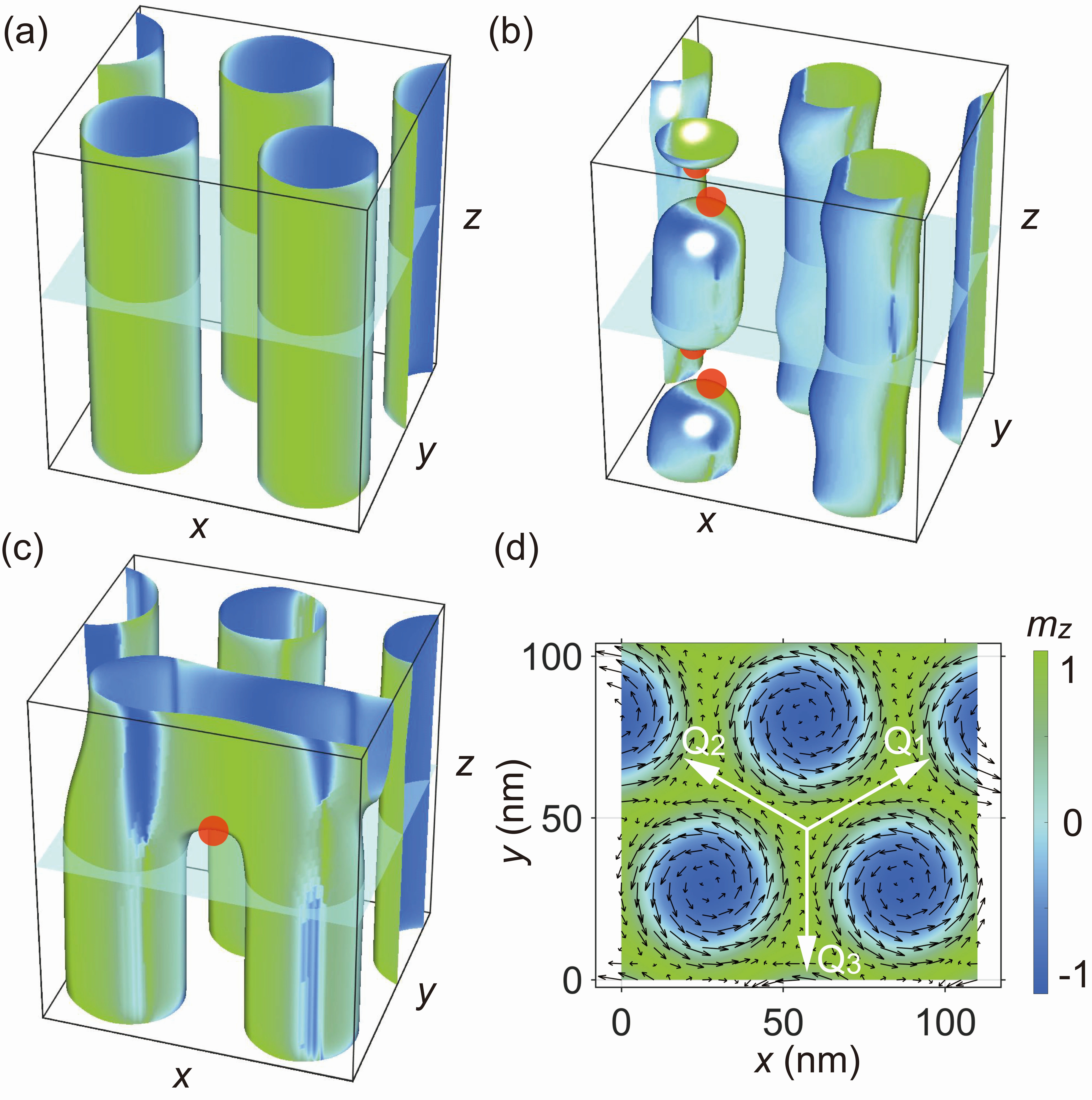}
\caption{3D simulation of the skyrmion lattice. 
(a) Isosurface visualization ($m_z = 0$) of a well-ordered 3D SkX phase. During the relaxation process, topological defects (red dots) emerge that break the local skyrmion strings, as shown in (b) and (c). 
(d) Shared 2D skyrmion plane, cut from the transparent slices in (a-c). The three $\mathbf{Q}_i$ wavevectors are shown.
}
\label{fig_4}
\end{figure}

As shown in Fig.\ 3(a-c),  both relaxation curves follow the same line shape that is described by an exponentially decaying profile, $I(t) = I_0 e^{-t/\tau}$, where $I = |\mathbf{M}|^2$, $I_0$ is the equilibrium peak intensity at $t_d<0$, and $\tau$ measures the characteristic timescale of the relaxation that is related to the intrinsic energy scale \cite{Tokura_MnSi_quench_Natphys_16,Pf_FeCoSi_LTEM_SciAdv_17}.   
Note that the other $\mathbf{Q}$ peaks have the same $t_d$-dependence.
The dynamic behavior is consistent across all three processes.
This consistency points at a universal relaxation mechanism.
The green lines in Fig.\ 3 show the exponential fit to the measured data with a lifetime of $\tau=0.12$, 0.19, and 0.19~s for the three processes, respectively.   
The difference in their exact value may be attributed to the differences in temperature between the phases, which govern the thermodynamics of the system \cite{Tokura_MnSi_quench_Natphys_16,Pf_FeCoSi_LTEM_SciAdv_17}.  
Unfortunately, due to the narrow SkX phase pocket, the $\tau(T)$ relationship over a wide temperature range becomes experimentally inaccessible. 
Nevertheless, the measured $\tau$ values belong to the same timescale of $\sim$0.2\,s, suggesting that the underlying mechanisms share the same origin.             
In order to gain a deeper understanding, 3D micromagnetic simulations were performed using a set of parameters for a typical chiral magnet \cite{Bobber_MCTR_PRL_21}.  
Figure 4(a) shows the skyrmion tube lattice of the equilibrium SkX phase in the OOP geometry, where the well-ordered 3D skyrmion crystal corresponds to the minimum global energy.
This leads to the superposition of the three modulation wavevectors, $\mathbf{Q}_i$, as shown by the cross-section plane in Fig.\ 4(d). 
Upon exciting the system with a slight tilt of the skyrmion tubes, and subsequently removing the titling field component, the system strives to relax back to a perfect skyrmion crystal. 
Nevertheless, during the relaxation process, local defects spontaneously nucleate. 
As indicated by the red dots in Fig.\ 4(b,c), emergent monopole/antimonopole structures emerge with certain randomness within or between the skyrmion strings.
They either cut the strings into separate sections [Fig.\ 4(b)], or merge the two neighboring strings [Fig.\ 4(c)], creating topological defects that are rather stable over the entire timescale accessible by micromagnetics. 
In either case, though the global energy is slightly higher, the system still orders as a rigid crystal, as the three $\mathbf{Q}$-wavevectors can be clearly identified, as shown by Fig.\ 4(d).  
Consequently, the RSM configuration remains the same for both [Fig.\ 4(a-c)], while only the amplitudes of the order parameter $\mathbf{M(Q)}$ are different. 
It is important to note that complementary measurements on the collinear field-polarized phase show no $t_d$ dependence \cite{SM}, while $t_d$ is independent of applied field in the conical phase \cite{SM}, making the slow relaxation process a unique and intrinsic feature of non-collinear orders.

Thus, the slow relaxation behavior observed in the pump-probe experiments essentially reflects the unwinding process of the as-nucleated topological defects. 
The decay curves in Fig.\ 3 can be interpreted in the following way.  
First, near $t_d=0$, the spontaneous nucleation of local defects render the magnetic crystal no longer to be a perfect spin lattice, hence a sudden decrease of the structure factor is observed.   
Such defects are local kinks (both structurally and energetically), therefore requiring significantly prolonged times to be dissipated. 
Consequently, the system slowly improves to a well-ordered magnetic crystal, recovering $|\mathbf{M(Q)|^2}$ to the same value as the $t_d<0$ state. 
Therefore, the measured $\tau$ essentially describes the lifetime of the emergent 3D topological defects. 
In summary, our study revealed a fundamentally different slow relaxation mechanism for non-collinear magnetic states beyond the micromagnetic framework.  
This slow relaxation process is due to the formation and dissipation of local topological defects.
The real-time unwinding process was quantitatively characterized using REXS, and the lifetimes of the defects in the various states were experimentally determined.
Importantly, the conical and SkX phases exhibit very similar relaxation profiles, indicating that their defect-governed dynamics is universal for their incommensurate non-collinear magnetic orders. 
The presence of topological defects not only plays a critical role in governing the dynamic behavior of these systems but also has profound implications for the design and performance of spintronic devices, which rely on rapid magnetic state changes. 
Therefore, understanding and controlling the formation and dissipation of topological defects are essential for optimizing the performance of future spintronic technologies.\\

\noindent \textbf{Acknowledgments}
This work was supported by the National Key R\&D Program of China under contract numbers 2022YFA1403602 and 2020YFA0309400, the Science and Technology Commission of the Shanghai Municipality (21JC1405100), the National Natural Science Foundation of China (Grant nos.\ 12074257 and 12241406), and the Double First-Class Initiative Fund of ShanghaiTech University.       
Diamond Light Source is acknowledged for beamtime on beamline I10 under proposal numbers MM34423-1 and MM34827-1, 
Prof.\ S.\ Giblin (Cardiff University) for provision of the magnetic coils, and J.\ Bollard and E.\ Arnold for their assistance during one of the beamtimes.

\hfill

\noindent

\noindent {\footnotesize{{$^\ast$These authors contributed equally to the work.}} \\
\noindent {\footnotesize{shilei.zhang@shanghaitech.edu.cn}}\\

%\bibliography{Skyrmion_People_v2a} 

%

\end{document}